**Electronic Properties of Strained Si/Ge Core-Shell Nanowires**

**Xihong Peng, [1*] Paul Logan[2]**

[1]Dept. of Applied Sciences and Mathematics, Arizona State University, Mesa, AZ 85212.

[2]Dept. of Physics, Arizona State University, Tempe, AZ 85287

## ABSTRACT

We investigated the electronic properties of strained Si/Ge core-shell nanowires along the [110] direction using first principles calculations based on density-functional theory. The diameter of the studied core-shell wire is up to 5 nm. We found the band gap of the core-shell wire is smaller than that of both pure Si and Ge wires with the same diameter. This reduced band gap is ascribed to the intrinsic strain between Ge and Si layers, which partially counters the quantum confinement effect. The external strain is further applied to the nanowires for tuning the band structure and band gap. By applying sufficient tensile strain, we found the band gap of Si-core/Ge-shell nanowire with diameter larger than ~3 nm experiences a transition from direct to indirect gap.

Strain engineering electronic properties of nanostructures has continuously attracted extensive interests. The Ge/Si system has a long history for introducing strain due to the large lattice mismatch up to 4%. A well known example is the Ge self-assembly pseudomorphic nanodots on Si substrates. Various nanostructures such as springs and nanotubes were also produced through the bilayer Si/Ge structure, which possesses unique electronic and optical properties.[1] Recently, Si/Ge core-shell nanowires have become particularly interesting both experimentally and theoretically,[2-7] due to their potential applications in electronic and optical devices. Compared to pure Si and Ge wires, the core-shell structure has some superior properties. For instance, a better conductance and higher mobility of charge carriers can be obtained, due to the band offsets in the core-shell nanowires.[8]

Si/Ge core-shell wires possess an intrinsic strain due to the lattice mismatch between Si and Ge. This intrinsic strain has been much less studied in the literature, although the strain has been demonstrated to critically affect the electronic properties of varies nanostructures.[3, 9-11] In this work, we investigated the strain effects on the band structure of Si/Ge core-shell wires along [110] direction with diameters up to 5 nm, using first principles calculations. We found the intrinsic strain has dramatically countered the quantum confinement effect and reduced the band gap of the core/shell nanowires. The band structure of the core/shell wires can be further notably modulated by an external uniaxial strain.

The density-functional theory (DFT)[12] calculations were performed using the codeVASP.[13] The DFT local density approximation (LDA) was applied. In detail, we used a pseudo-potential plane wave approach with kinetic energy cutoff of 200.0 eV with the key tag PREC = Accurate. The core electrons are described using Vanderbilt pseudo-potentials.[14] The reciprocal space is sampled at 1x1x4 using Monkhorst Pack meshes. We include 21 k-points in the band structure calculation along Γ to X. The dangling bonds on the surface are saturated by hydrogen. For Si-core/Ge-shell nanowires, the core is always consisted of 30 Si atoms with a diameter 1.5 nm, and the thickness of the Ge shell is varied with the wire size. Similarly, for Ge-core/Si-shell wires, the core contains 30 Ge atoms. The initial lattice constants in Si-core/Ge-shell and Ge-core/Si-shell nanowires are set to be 0.3977 nm and 0.3862 nm, taken from bulk Ge 0.5625 nm and bulk Si, 0.5461 nm (*i.e.*, $a_{initial[110]} = a_{bulk} / \sqrt{2}$), respectively. The lateral length of the simulation cell is chosen so that the distance between the wire and its replica (due to periodic boundary conditions) is more than 1.0 nm to avoid the interactions between the wire and its replica. The wire is then optimized through total energy minimization till the force acting on atoms is less than 0.02 eV/Å. The band gap of a wire is defined by the energy difference between the conduction band edge (CBE) and the valence band edge (VBE). Once the optimized wire is obtained, we applied uniaxial tensile/compressive (i.e. positive/negative) strain by scaling the axial lattice constant. For each strained wire, the lateral $x$ and $y$ coordinates are further relaxed. Our study showed the band structure of the wire is affected significantly by strain.

First we report the geometrical structure of the relaxed Si/Ge core-shell nanowires. The lattice constant $a_{bulk}$ in bulk Si and Ge are 0.5461 nm and 0.5625 nm, respectively, based on the simulation parameters mentioned above. In order to optimize the axial lattice constant for a given wire, we performed a series of calculations on the total energy with different lattice constants. The optimized lattice constant $a_{optimized}$ was obtained by parabolic fitting the graph of total energy versus lattice. The axial stress in the wire was minimal for the optimized constant. The results of $a_{optimized}$ are presented in Fig. 1(a). In Si-core/Ge-shell wires, the lattice constant is increasing with the diameter of the wire, from 0.3917 nm for the 2.5 nm wire to 0.3944 nm for the 4.7 nm wire. In addition, the lattice constants are smaller than 0.3977 nm (from bulk Ge), but larger than 0.3862 nm (from bulk Si). These results are expected since a larger wire has more Ge atoms in the shell. On the other hand, the lattice constant of Ge-core/Si-shell wires is reduced with size, from 0.3985 nm to 0.3900 nm. More interestingly, the lattice constant for the smallest wire with the diameter 2.5 nm is even larger than 0.3977 nm from bulk Ge. This is consistent with the findings in pure Si and Ge nanowires[10, 11] that the smaller wires are expanded. Figure 1(b) shows the intrinsic strain produced in the core-shell wires. We can see that the Si composition is in tensile strain (i.e. positive strain), while the Ge composition is in compressive strain (i. e. negative strain).

In Fig. 2, the band structures of Si/Ge core-shell nanowires with varied diameters clearly demonstrates a direct band gap at Γ, consistent with previous study.[6,10,11] In addition, when examining the band edges, we noticed that for Si-core/Ge-shell wires (Figs. 2a – 2c), the valence and the lower occupied bands are generally close to each other, while the conduction and the higher unoccupied bands are considerably discrete. However, for Ge-core wires (Figs. 2d – 2f), the energy spacing in occupied bands are larger than that of conduction bands. According to the band lineup in Si/Ge core-shell nanowires,[6] the wave function of the valence band is mainly contributed by Ge atoms, while the orbital of the conduction band is primarily located in Si atoms. The discrete levels of the conduction bands of Si-core/Ge-shell wires in Figs. 2a – 2c implies Si atoms are more significantly quantum confined, consistent with the fact that they are in the core. On the other hand, the Ge atoms in Ge-core/Si-shell wires are confirmed to be quantum confined by the discrete valence bands in Figs. 2d – 2f.

In Fig. 3a, we report the DFT predicted band gaps for the wires. Although the accurate calculation of band gap requires advanced GW method[15-17] or quantum Monte Carlo calculations[18-20], we expect the DFT gaps in the present work can give a qualitative prediction in the size-dependency, as shown in ref. 21. We found the gap of the Si/Ge core-shell nanowire is increased with reducing size (Fig. 3a). This effect is primarily due to the quantum confinement effect.

More interestingly, we can see that the band gap of the core-shell nanowires is smaller than that of both pure Si and Ge nanowires with the same diameter (Fig. 3a). For example, the DFT gap for Si and Ge wires with the diameter 2.5 nm are 1.02 eV and 0.73 eV,[11, 10] respectively. However the gap for the Si/Ge core-shell wires are 0.58 eV(Ge-core) and 0.54 eV (Si-core), respectively, which are both smaller than that of Si and Ge wires. Similar trends are also observed for the larger wires. In order to understand this behavior, we took an example of a Si-core/Ge-shell nanowire with a diameter of 3.0 nm. The gap of this wire was reduced by 0.3 eV compared to the pure 3.0-nm-Ge-wire. From the lattice and intrinsic strain reported in Fig. 1, the Si-core experiences a 2.2% tensile strain and the Ge-shell is contracted with a 0.8% compressive strain , compared to bulk. Figs. 3b and 3c present the energies of the CBE and VBE in pure Si and Ge wires as a function of uniaxial strain.[10, 11]As reported,[6] the CBE of the core-shell nanowire is contributed by Si, while the VBE by Ge. In Fig. 3b, we found that the CBE of the Si wire with a diameter of 1.5 nm (i.e. the size of the core) is decreased by ~ 0.2 eV under a 2.2% tensile strain; while, in Fig. 3c, the change of the VBE in the Ge wire with a diameter of 3.0 nm is negligible under a 0.8% compressive strain. Therefore, we would expect the band gap in a Si-core/Ge-shell nanowrie with a diameter of 3.0 nm was reduced by 0.2 eV compared to the pure 3.0-nm-Ge wire. This reduction is close to the actual calculation (~ 0.3 eV), in Fig. 3a. Similar qualitative behaviors are also observed in the Ge-core/Si-shell nanowires.

From the above analysis, we can see that the reduce band gap is closely related to the intrinsic strain of the core/shell wires. Amato *at el.*[7] also observed a reduced band gap in Si/Ge nanowires which form explicit interface between Si and Ge regions. The authors explained the gap reduction using quantum confinement effects in the band edges. This quantum confinement effect may be able to explain their wires with diameters up to 1.6 nm. However, it is not able to explain our larger wires with diameters up to 5 nm, where the quantum confinement has been weak. We argue that the reduction of the gap in the core-shell wires is mainly due to the intrinsic strain in the Si/Ge composition.

We further studied the band structures of Si/Ge core-shell nanowires modulated by external uniaxial strain. In Fig. 4a, we compared the band structures of the Si-core/Ge-shell wire with a diameter of 3.0 nm, without and with strain. Solid lines are the band structure without strain; dashed lines are under 2.8% uniaxial tensile strain; dotted lines are under 2.2% compression. Generally, strain shows a dominant effect on the band near Γ - the energy levels are shifted evidently with strain (see the dashed oval). However, strain has negligible effects on the band far away from Γ - a minimal energy shift under strain (see the solid ovals). This strain effect on the band structure is understandable from the tight-binding model.[22, 10] the band energies of a wire is written as $E(k) = E_v - \beta - 2\gamma \cos(k_{//} a/2)$, where $E_v$ is the energy of atomic orbitals, $k_{//}$ is the wave vector, $\beta$ is a small energy correction near the nucleus site, $\gamma$ (the overlap integral) is an energy correction dependent on the orbital overlap between two neighboring atoms, and $a$ is the lattice constant. For Γ ($k_{//} = 0$), the energy is $E(\Gamma) = E_v - \beta - 2\gamma$. For X ($k_{//} = \pi / a$), the energy is $E(X) = E_v - \beta$. Applying strain to a wire, the bond length of Si-Ge will be changed. Thus we expect a prominent modification of the $\gamma$ value, while a negligible variation of $\beta$ due to its local nature. Referring to the above formulae of $E(\Gamma)$ and $E(X)$, strain will bring a more pronounced effect in the energy at $\Gamma$, compared to other K points.

We also noticed that under large extensile strain, such as 2.8%, the VBE for the Si-core wire with diameter 3.0 nm experiences a significant change – the top of the band is no longer located at Γ, implying an indirect band gap. An enlarged graph of the valence band near Γ without strain and under 2.8% extension is presented in Fig. 4b. The real valence band edge is determined by the two states labeled as $v_0$ at Γ and $v_1$ near K = 0.05 $(2\pi/a)$. Without strain (refer to the black solid line), $v_0$ is higher than $v_1$, the wire demonstrates a direct band gap. Under 2.8% uniaxial extension, the shift-down of $v_0$ is more significant than that of $v_1$, so that $v_1$ becomes the valence band edge, resulting in an indirect band gap. This direct-to-indirect-band-gap transition also happens for the larger Si-core wires with diameters 3.7 nm and 4.7 nm under tensile strain. The tensile strain induced direct-to-indirect gap transition was also previously found in Si nanowires and Ge nanowires. [9, 11] However, for Ge-core/Si-shell nanowires, we did not find similar transition under the amount of the strain being applied in the present work.

In summary, we found that (1) due to the intrinsic strain in the Si/Ge core-shell nanowires, the direct band gap is smaller than that in pure Si and Ge wires given the same diameter; (2) the Si-core/Ge-shell nanowire may be tuned to indirect band gap by applying sufficient tensile strain.

This work is supported by the Research Initiative Fund from Arizona State University. We are thankful to ASU-Saguaro and NCSA for computational resources. Jeff Drucker and Fu Tang are acknowledged for helpful discussions.

* Corresponding author, xihong.peng@asu.edu

## REFERENCE:

1. M. Huang, C. Boone, M. Roberts, D. E. Savage, M. G. Lagally, N. Shaji, H. Qin, R. Blick, J. A. Nairn, F. Liu, Adv. Mater. **17**, 2860 (2005).
2. Y. He, Y. Zhao, C. Fan. X. Liu, R. Han, Front. Electr. Electron. Eng. China **4**, 342 (2009).
3. I. A. Goldthorpe, A. F. Marshall, P. C. Mclntyre, Nano Lett. 9, 3715 (2009) ; Nano Lett. **8**, 4081 (2008).
4. T. E. Trammell, X. Zhang, Y. Li, L.-Q. Chen, E. C. Dickey, J. Crystal Growth **310**, 3084 (2008).
5. R. N. Musin, X. –Q. Wang, Phys. Rev. B **74**, 165308 (2006).
6. L. Yang, R. N. Musin, X. –Q.Wang, M. Y. Chou, Phys. Rev. B **77**, 195325 (2008).
7. M. Amato, M. Palummo, S. Ossicini, Phys. Rev. B **79**, 201302 (2009).
8. J. Xiang, W. Lu, Y. Hu, H. Yan, C. M. Lieber, Nature **441**, 489 (2006).
9. P. W. Leu, A. Svizhenko, K. Cho, Phys. Rev. B **77**, 235305 (2008).
10. P. Logan, X.-H. Peng, Phys. Rev. B **80**, 115322 (2009).
11. X. -H. Peng, A. Alizadeh, S. K. Kumar, and S. K. Nayak, Int. J. of Applied Mechanics **1**, 483 (2009).
12. P. Hohenberg and W. Kohn, Phys. Rev. **136**, B864 (1964); W. Kohn and L.J. Sham, Phys. Rev. **140**, A1133 (1965);
13. G. Kresse and J. Furthmüller, Phys. Rev. B **54**, 11169 (1996); G. Kresse, J. Furthmuller, Comput. Mater. Sci. **6**, 15 (1996).
14. D. Vanderbilt, Phys. Rev. B **41**, 7892 (1990).
15. L. Hedin, Phys. Rev. **139**, A796 (1965).

16. S. V. Faleev, M. V. Schilfgaarde, and T. Kotani, Phys. Rev. Lett. **93**, 126406 (2004).

17. F. Bruneval, F. Sottile, V. Olevano, R. D. Sole, and L. Reining, Phys. Rev. Lett. **94**, 186402 (2005).

18. A. Puzder, A. J. Williamson, J. C. Grossman, and G. Galli, Phys. Rev. Lett. **88**, 097401 (2002); Mat. Sci. & Eng. B **96**, 80 (2002); J. Chem. Phys. **117**, 6721 (2002).

19. J. Williamson, J. C. Grossman, R. Q. Hood, A. Puzder, and G. Galli, Phys. Rev. Lett. **89**, 196803 (2002).

20. X.-Y. Zhao, C. M. Wei, L. Yang and M. Y. Chou, Phys. Rev. Lett. **92**, 236805 (2004).

21. X.-H. Peng, S. Ganti, A. Alizadeh, P. Sharma, S. K. Kumar, S. K. Nayak, Phys. Rev. B **74**, 035339 (2006); X.-H. Peng, A. Alizadeh, N. Bhate, K. K. Varanasi, S. K. Kumar, and S. K. Nayak, J. Phys.: Condens. Matter **19**, 266212 (2007).

22. N. W. Ashcroft and N. D. Mermin, Solid State Physics,Thomson Learning, 1976; M. A. Omar, Elementary Solid State Physics, Addison-Wesley Publishing Company, MA, 1993.

## Figure captions

**Fig. 1. (a) The optimized axial lattice constant of Si/Ge core-shell nanowires as a function of the wire diameter. As a reference, the lattice constants along [110] direction derived from bulk Ge and Si are given by the dotted and dashed lines, respectively; (b) the intrinsic strain experienced by the Si and Ge composition under the optimized axial constant.**

**Fig. 2. The band structures of Si-core (a-c) and Ge-core (d-e) nanowires with varied diameter. The Fermi level of the wires is aligned at zero. The band edges are highlighted by the dashed oval. It shows that, the conduction and the upper unoccupied bands in Si-core wires are affected significantly by the quantum confinement, while the valence and the lower occupied bands in Ge-core wires are quantum confined.**

**Fig. 3 (a) The DFT gap of Si/Ge core-shell wires as a function of the wire diameter. The band gap of the corresponding pure Si and Ge wires are also plotted for comparison. (b) The energies of the CBE and VBE in the pure Si nanowire with a diameter 1.5 nm versus. (c) The energies of the band edges in the pure Ge nanowire with a diameter 3.0 nm versus strain.**

**Fig. 4 (a) The band structures of Si-core/Ge-shell wire with a diameter of 3.0 nm, without and under strain. Solid lines are the band structure without strain; dashed lines are under 2.8% tensile strain; dotted lines are under 2.2% compressive strain. (b) The enlarged graph of the VBE without (solid line) and under 2.8% extension (dashed curve). The inset is a magnified image of the rectangle.**

(a)

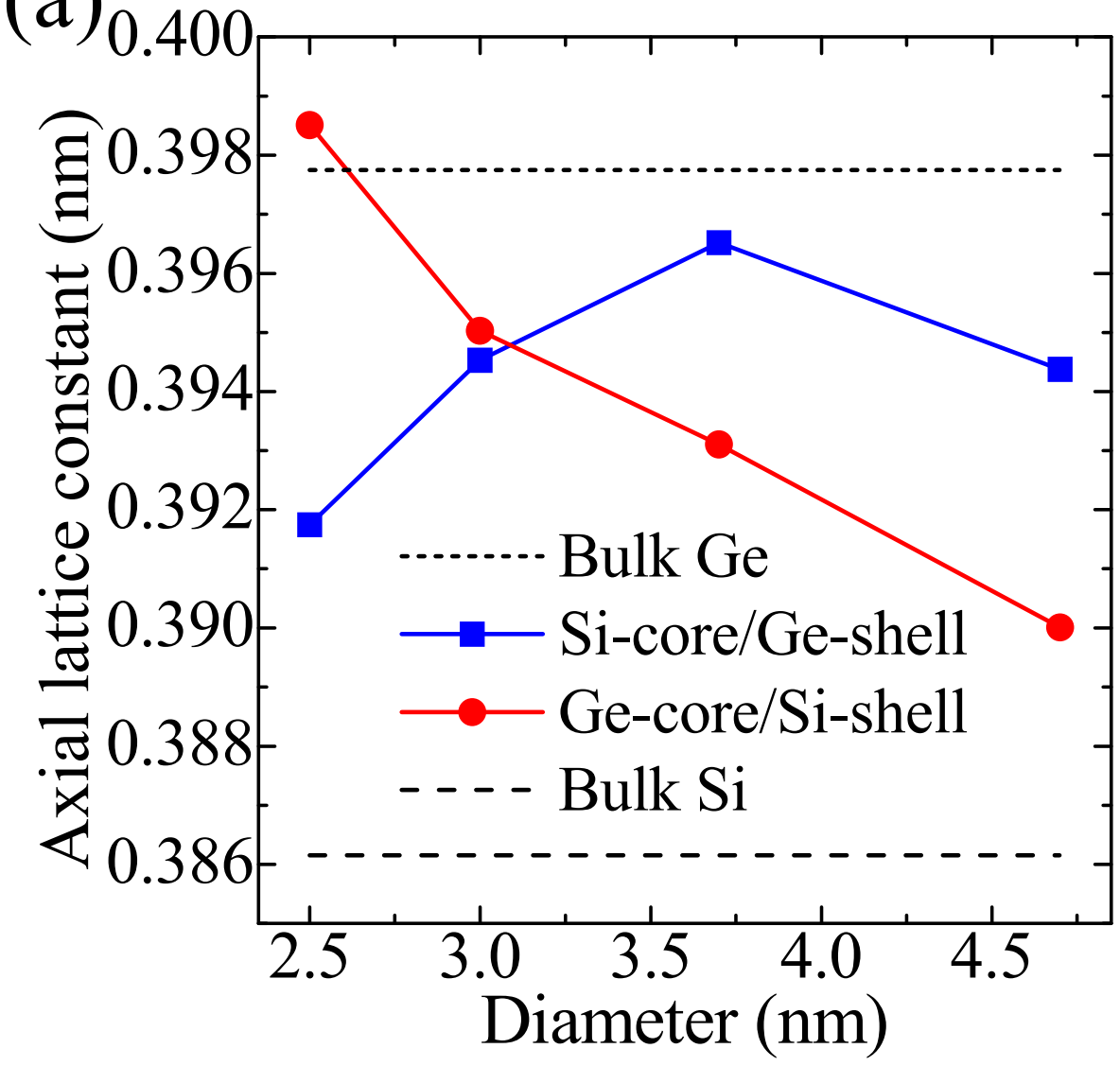

(b)

| | Diameter (nm) | Strain to core (%) | Strain to shell (%) |
|---|---|---|---|
| Si-core /Ge-shell | 2.5 | 1.5 | -1.5 |
| | 3.0 | 2.2 | -0.8 |
| | 3.7 | 2.7 | -0.3 |
| | 4.7 | 2.1 | -0.8 |
| Ge-core /Si-shell | 2.5 | 0.2 | 3.2 |
| | 3.0 | -0.7 | 2.3 |
| | 3.7 | -1.2 | 1.8 |
| | 4.7 | -1.9 | 1.0 |

**FIG. 1. Peng *et al*.**

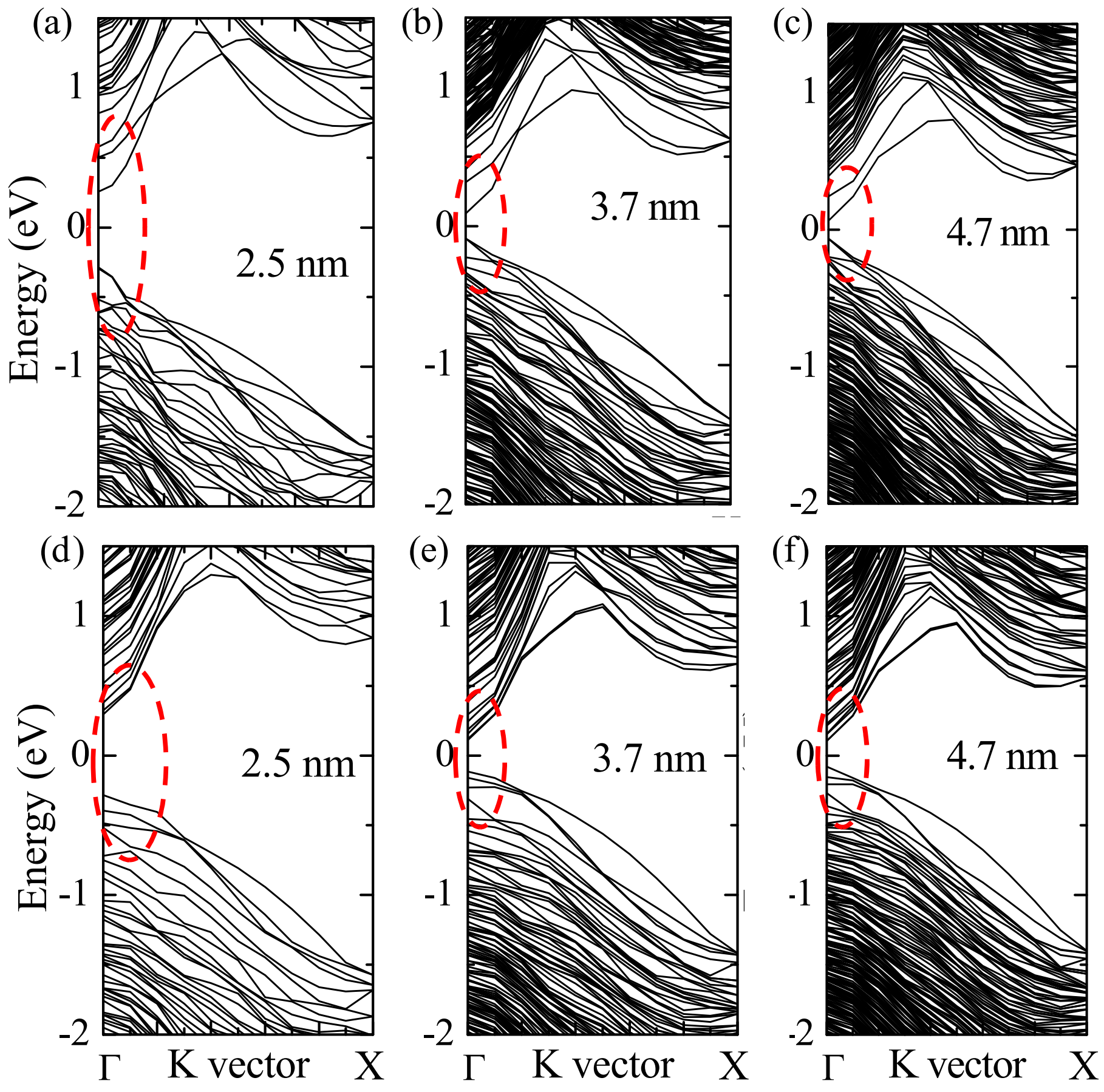

**FIG. 2, Peng *et al*.**

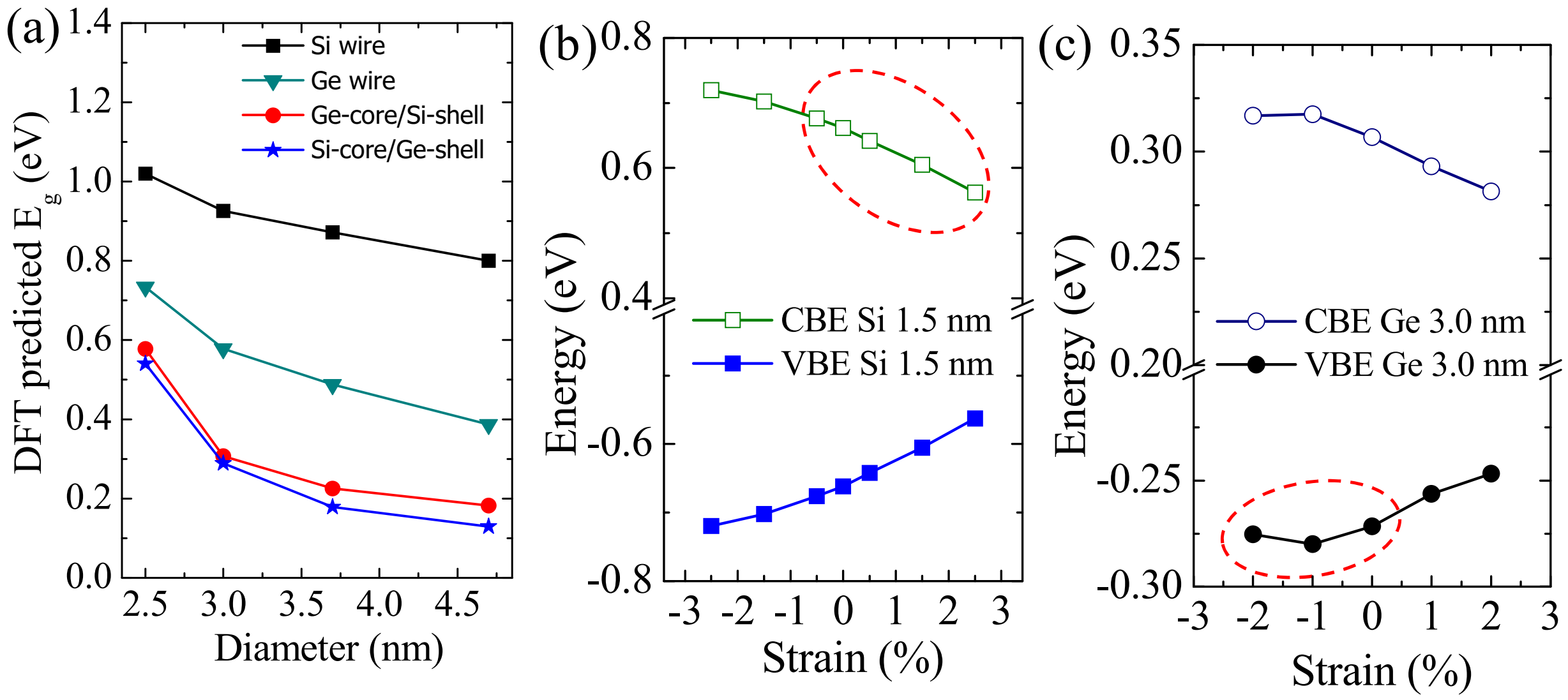

FIG. 3, **Peng *et al*.**

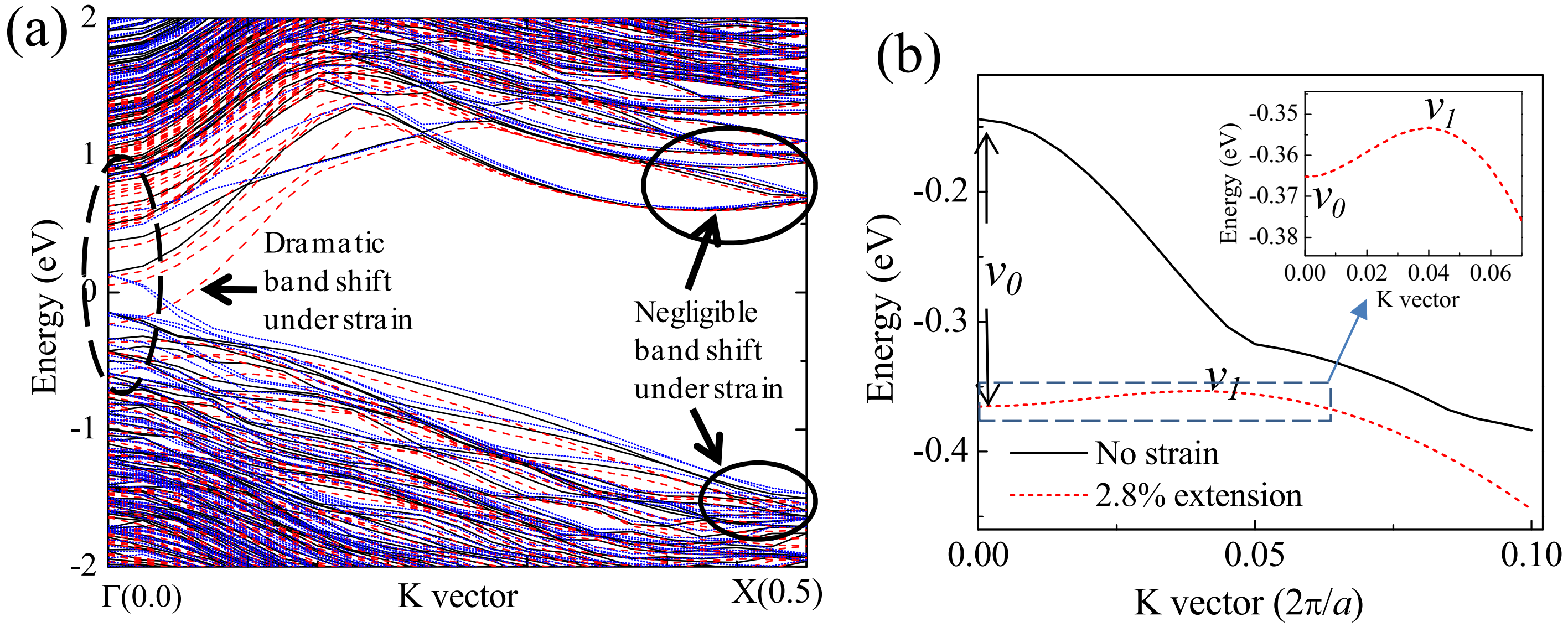

**FIG. 4,** Peng *et al*.